\documentclass[letterpaper]{article} 
\usepackage{aaai24}
\usepackage{times}  
\usepackage{helvet}  
\usepackage{courier}  
\usepackage[hyphens]{url}  
\usepackage{graphicx} 
\usepackage{amssymb}
\usepackage{amsmath}
\usepackage{natbib}  
\usepackage{caption} 
\usepackage{algorithm}
\usepackage{algorithmic}
\usepackage{newfloat}
\usepackage{listings}
\DeclareCaptionStyle{ruled}{labelfont=normalfont,labelsep=colon,strut=off} 
\floatstyle{ruled}
\newfloat{listing}{tb}{lst}{}
\floatname{listing}{Listing}
\title{AffectEcho: Deep Neural Architecture for Textless Speaker independent and Language Independent Modelling of affect and emotions in spoken speech samples}

\title{\textit{AffectEcho}: Speaker Independent and Language-Agnostic Emotion and Affect \\Transfer for Speech Synthesis}

\author {
    Hrishikesh Viswanath\textsuperscript{\rm 1},
    Aneesh Bhattacharya\textsuperscript{\rm 1},
    Pascal Jutras-Dube\textsuperscript{\rm 1},
    Prerit Gupta\textsuperscript{\rm 1},
    Mridu Prashanth\textsuperscript{\rm 1},
    Yashvardhan Khaitan\textsuperscript{\rm 1},
    Aniket Bera\textsuperscript{\rm 1}
}
\affiliations {
    \textsuperscript{\rm 1}Department of Computer Science, Purdue University, West Lafayette, IN, USA\\
    hviswan@purdue.edu, bhatta95@purdue.edu, pjutrasd@purdue.edu, gupta596@purdue, mprasha@purdue.edu, ykhaita@purdue.edu,aniketbera@purdue.edu
}

\usepackage{bibentry}

\begin{document}

\maketitle

\begin{abstract}
Affect is an emotional characteristic encompassing valence, arousal, and intensity, and is a crucial attribute for enabling authentic conversations. While existing text-to-speech (TTS) and speech-to-speech systems rely on strength embedding vectors and global style tokens to capture emotions, these models represent emotions as a component of style or represent them in discrete categories. We propose AffectEcho, an emotion translation model, that uses a Vector Quantized codebook to model emotions within a quantized space featuring five levels of affect intensity to capture complex nuances and subtle differences in the same emotion. The quantized emotional embeddings are implicitly derived from spoken speech samples, eliminating the need for one-hot vectors or explicit strength embeddings. Experimental results demonstrate the effectiveness of our approach in controlling the emotions of generated speech while preserving identity, style, and emotional cadence unique to each speaker. We showcase the language-independent emotion modeling capability of the quantized emotional embeddings learned from a bilingual (English and Chinese) speech corpus with an emotion transfer task from a reference speech to a target speech. We achieve state-of-art results on both qualitative and quantitative metrics. 
\end{abstract}

\section{Introduction}

Expressions of similar emotions can exhibit subtle variations across different languages, and the manifestation of these emotions can also differ among individuals \cite{jackson2019emotion}. The integration of these nuanced emotional qualities into conversational AI systems holds the potential to enhance human-AI interactions, enabling more realistic and engaging dialogues \cite{martinez2005emotions}. However, accurately modeling emotions in a language-independent setting presents a significant challenge due to the need to group similar emotions with subtle differences together \cite{van2023modelling}. The complexity arises from the fact that individuals simultaneously express a multitude of emotions with varying intensities during speech, making a one-to-one mapping of emotions impractical \cite{cowen2019primacy}. Instead, we devise a new approach that involves clustering and quantizing similar emotions into embeddings which can guide generative models.

Numerous deep learning approaches have been developed for tasks such as text-to-speech synthesis (TTS) \cite{ren2019fastspeech} and speech style transfer, achieving impressive results in generating high-quality speech \cite{zhou2021seen}. Recent advancements have also focused on synthesizing emotional speech using methods such as one-hot vectors or strength vectors \cite{zhou2022mixed}, diffusion-based techniques to control emotional intensity \cite{guo2023emodiff}, and generating speech with mixed emotions \cite{zhou2022mixed}. However, in conversational AI systems, it is impractical to provide explicit strength vectors for every response, therefore necessitating models that can comprehend the speaker's emotions automatically and respond appropriately. Moreover, it is crucial that the generated response avoids aggravating the emotional state of the human speaker interacting with the AI agent \cite{carolus2021towards}. To address this challenge, as a first step, we present a deep learning-based pipeline that effectively captures the speaker's emotion and faithfully mirrors the same emotion while preserving key attributes such as identity, accent, linguistic content, intonation, and cadence.

Several studies have explored the unsupervised modeling of speaking styles with style tokens, which serve as embeddings capable of capturing various characteristics such as speed, speaking style, prosody, and speaker identity \cite{wang2018style}. While these style token embeddings can be trained to capture emotions, it is essential to decouple style and emotion to enhance control over speech generation \cite{li2021controllable}. In the context of dialogue systems, conversational agents, exemplified by popular virtual assistants like Siri, have pre-defined identities, accents, and prosody. Integrating style transfer models into such agents would be inefficient as they are designed to maintain a consistent identity throughout interactions with humans. 

In recent years, significant advancements have been made in text-to-speech (TTS) systems, resulting in the development of state-of-the-art models capable of generating highly realistic speech from text input. Our proposed model can effectively complement these advanced TTS pipelines by acting as a post-processing step, enabling the integration of appropriate emotions into the synthesized speech prior to its delivery to the individual engaged in interaction with the conversational agent. By leveraging the existing strengths of TTS systems in generating accurate speech content, our model focuses on infusing the desired emotional nuances, thereby enhancing the overall quality and expressiveness of the generated speech.

The main contributions of our research work can be summarized as follows:
\begin{itemize}
\item We introduce a methodology using a Vector Quantized codebook model to learn meaningful affect representations from speech, capturing variations in valence, arousal and dominance within an emotion, while, eliminating the need for one-hot representations and explicit strength embeddings of emotions.

\item We design \textbf{AffectEcho}, an emotional speech conversion model conditioned on the quantized emotional embeddings. Our method disentangles emotion from style and linguistic content, facilitating cross-language emotion transfer and offering enhanced flexibility and controllability. Furthermore, we use spectral convolution blocks or neural operator blocks to better learn the acoustic features in the spectral domain. 

\item Through quantitative and qualitative experiments, we demonstrate that \textbf{AffectEcho} can successfully transfer emotion between speakers while preserving their unique characteristics such as speaking style, linguistic content, and vocal characteristics. 

\end{itemize}

\section{Related Work}
Recent advancements in text-to-speech architectures have led to innovative approaches to emotional speech synthesis. 
\citet{zhou2022mixed} propose a sequence-to-sequence architecture that enables emotional manipulation in text-to-speech models. 
While their approach utilizes editable strength embeddings to guide the generator, it focuses on discrete emotions and overlooks other affective features like valence, dominance, and arousal.
Another avenue explored by \citet{guo2023msmc} involves a multi-stage codebook for text-to-speech conversion. 
Their approach employs VQ-VAE \cite{van2017vqvae} to quantize acoustic features of the mel-spectrogram and subsequently reconstruct the target audio. 
In contrast, many of the recent approaches in text-to-speech leverage diffusion generative models \cite{dickstein2015diffusion}.
Notable among these is EmoDiff \cite{guo2023emodiff}, which focuses on intensity-controllable diffusion-based text-to-speech modeling. 
It uses weighted emotion labels during sampling to generate speech samples with the desired emotions at the potential cost of reduced expressiveness in other affective dimensions in the condition vector. 
EmoMix \cite{tang2023emomix} conditions the diffusion training process on emotional embeddings derived from a pre-trained speech emotion recognition model.
NaturalSpeech2 \cite{shen2023naturalspeech} is a diffusion-based text-to-speech model that retains speaker identity and can perform speech enhancement. 
These models combine emotion and style, presenting a challenge in isolating and controlling emotion without compromising other features of the speaker's identity.

The use of class-conditioned StarGAN has been explored by \citet{rizos2020stargan} for emotion conversion, where emotions are represented as three mutually exclusive classes: angry, happy, and sad. 
\citet{luo2019neutral} introduce a GAN-based model employing continuous wavelet transform for neutral to emotional voice conversion, utilizing a Variational Autoencoder to extract the emotional information. 
Their approach also utilizes only three emotions.
\citet{kameoka2018starganvc} design the StarGAN-VC architecture for many-to-many voice conversion, later extended by \citet{das2023stargan} with StarGAN-VC++ to include style conversion to change the speaker identity while retaining the content of the speech. 
Similarly, \citet{shah2023nonparallel} propose emotion conversion using StarGAN-VC, incorporating dual encoders to learn speaker and emotion style embeddings separately. However, their model reported a low classification accuracy. 
\citet{meftah2023english} perform a comprehensive analysis of StarGAN-based models for emotional voice conversion and conclude that the efficiency in converting multi-emotions to multi-speakers was not as high as the efficiency in voice conversion for multi-speakers.

To address the lack of comprehensive affect-based emotion modeling in existing works and to decouple emotion from speaker style, we present AffectEcho, a textless speech affect transfer model. 
In our work, we showcase that emotional voice conversion benefits from the rich emotion representations learned via the proposed Vector Quantized codebook model, and that affect can be decoupled from gender, speaker, and language for better control over the generated speech. 

\section{AffectEcho Architecture}
\begin{figure}[h!]
    \centering
    \includegraphics[width=8cm]{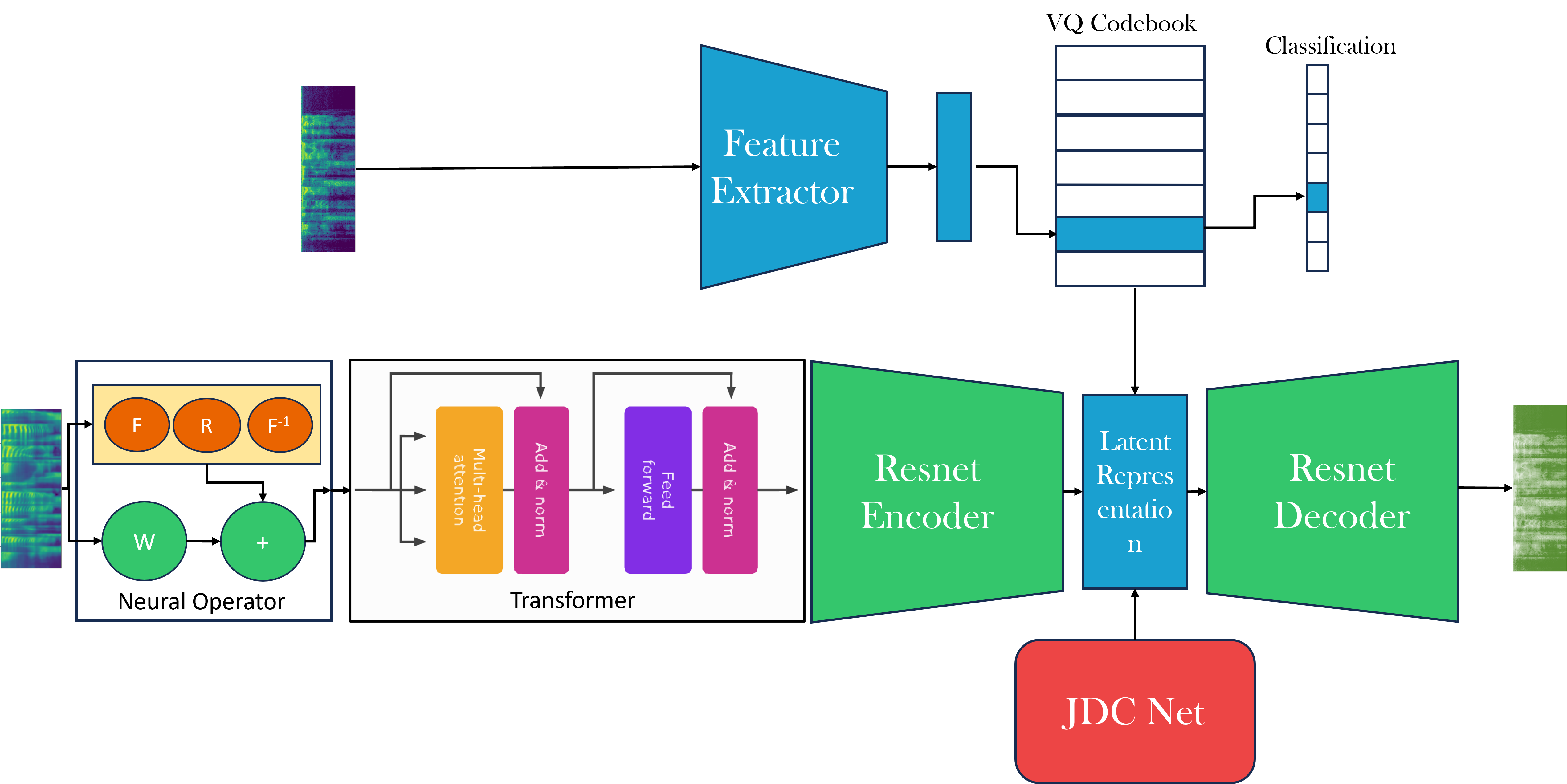}
    \caption{\textit{The overall architecture of \textbf{AffectEcho} Model. The figure on the top is the VQ classifier, which processes the input audio and maps it to the codebook. In the bottom row, the mel-spectrogram passes through neural operator blocks, transformer blocks and the encoder-decoder block.}}
    \label{fig:architecture}
\end{figure}
The \textbf{AffectEcho} model comprises two essential components. The first component is the emotion classifier responsible for generating the emotion embedding from the reference audio. This emotion embedding serves as a condition for the second component, the speech generator model. The speech generator takes the mel-spectrogram of the input speech as its primary input and leverages the emotion embedding to synthesize output speech with the desired affective qualities. Figure \ref{fig:architecture} summarizes the two components of the AffectEcho model. 

\subsection{Modeling Emotion Space using a Vector Quantized Classifier}
To generate the emotion embeddings, we use a vector quantized (VQ) classifier model. Recognizing the complexity of emotions and the limitations of binning them into discrete categories, we adopt a two-stage representation for these embeddings. Initially, we classify the dataset into five main categories, namely, angry, happy, neutral, sad, and surprised, each representing the dominant emotion exhibited in the speech sample. Subsequently, in the second stage, we leverage a vector quantized codebook to further amplify the emotion space, achieving five levels of nuanced representations per emotion. 

The VQ codebook defines the embedding space of $25$ categorical emotion vectors with $64$ affective features each. Each categorical vector $e_i \in \mathbb{R}^{64}$, $i \in \{1, \ldots, 25\}$ is manually associated with a one-hot encoded dominant emotion $p(e_i)$. 

Our VQ classifier first outputs an embedding vector $z(x) = e \in \mathbb{R}^{64}$, where $x$ is an input of speech features. The emotion category $z$ is then chosen by a nearest-neighbor look-up in the VQ codebook using cosine similarity $S_c(\cdot,\cdot)$ between the output vector and the $25$ categorical vectors of the codebook as shown in equation \ref{eq:codebook}: 
\begin{equation}
    \label{eq:codebook}
    q(z = k\mid x) = \begin{cases}
                1 & \text{if } k = \arg\max_{i \in [25]} S_c(z(x), e_i)\\
                0 & \text{otherwise}.\\
                \end{cases}
\end{equation}
In the loss, we use a categorical cross-entropy term between the input emotion label  and the emotion label associated with $z$, as indicated by equation \ref{eq:cce}:
\begin{equation}
    \label{eq:cce}
    L_{ce} = -\frac{1}{25}\sum_{i=1}^{25} p(e)_i \log(q(z \mid x)_i).
\end{equation}
%
%

Similar to what is described in the work of \citet{van2017vqvae}, we use $l_2$ error to move the embedding vectors $e_i$ towards $z(x)$ and a commitment loss to ensure the model commits to the embedding space
\begin{equation}
    \label{eq:commitment}
    L_{cl} = ||\text{sg}[z(x)] - e||_2^2 + \beta ||z(x) - \text{sg}[e]||_2^2.
\end{equation}
In equation \ref{eq:commitment}, $sg$ refers to the stop gradient operator, and $\beta$ is set to $0.25$ for all the experiments. 

The overall loss of the VQ-classification model is given by equation \ref{eq:classloss}:
\begin{equation}
    \label{eq:classloss}
    L_{vq} = L_{ce} + \alpha L_{cl}
\end{equation}
where $\alpha$ is set to $0.01$.
\subsubsection{Emotional Speech Generator}
The generator model $G$ is trained to transfer the emotion from a reference speech $y$ to a neutral speech $x$. It is structured with several key components: a spectral convolution block (neural operator) and a transformer block, followed by a ResNet encoder and decoder, similar to the generator from StarGAN-VC architectures. The generator maps the mel-spectrogram of the input speech to a new mel-spectrogram featuring the same speaking style but incorporating the affective features of the reference speech. 

To this end, two conditions are provided to guide the model during the decoding step. The first condition involves a pretrained JDC style encoder model \cite{kum2019joint}, which extracts fundamental frequencies from the input mel-spectrogram. It indicates to the model the desired speaking style for the generated speech. The second condition is the emotion embedding vector derived from the reference speech. It encourages the generator to generate speech samples exhibiting the same emotion as the reference speech.

To train the speech generator model, we use four loss functions. 
The first one is the reconstruction loss, which computes the $l_1$ loss between the generated and the target mel-spectrogram, given by equation \ref{eq:req}
\begin{equation}
    \label{eq:req}
    L_{rc} = || G(x) - y ||_1
\end{equation}
The second loss is the spectral convergence loss, which is the normalized Euclidean distance between the spectrograms denoted by equation \ref{eq:spc}:
\begin{equation}
    \label{eq:spc}
    L_{sc} = \frac{||G(x)||_2 - ||y||_2}{||G(x)||_2}.
\end{equation}
The third loss that we define to maintain the vocal cadence of the speaker called the pitch flow loss, which minimizes the difference in pitch with time, as shown in equation \ref{eq:pf}:
\begin{equation}
    \label{eq:pf}
    L_{pf} =  \left|\left| \sum_{t=1}^T(G(x)_{t+1} - G(x)_t) - \sum_{t=1}^T(y_{t+1} - y_t) \right|\right|_2.
\end{equation}
Lastly, we define speech emotion loss, which ensures that the target speech has the same emotion as the reference speech. We use the VQ classifier model to generate this loss indicated in equation \ref{eq:erl}:
\begin{equation}
    \label{eq:erl}
    L_{ser} = -\frac{1}{25}\sum_{i=1}^{25} q(z \mid y)_i \log(q(z \mid x)_i).
\end{equation}
The overall loss function is given by equation \ref{eq:loss}:
\begin{equation}
    \label{eq:loss}
    L_{gen} = L_{rc} + \alpha_1 L_{sc} + \alpha_2 L_{pf} + \alpha_3 L_{ser}.
\end{equation}

\subsubsection{Dataset}
The VQ-classifier was first trained on the MEAD dataset \cite{wang2020mead}. The classifier was fine-tuned on bilingual ESD dataset \cite{zhou2021seen}, containing the same utterances in different emotions, uttered in both English and Chinese. 
\subsubsection{Training} The VQ-classifier was trained on MEAD and ESD dataset for 20 epochs, with a 70-20-10 split. The Generator model was trained for 200 epochs and a batch size of 10. This model was trained on the ESD dataset (70-20-10 split) on Nvidia A30 GPU. 

\section{Evaluations}

\subsection{Quantitative Metrics}

\subsubsection{Target Speech Emotion Recognition}
In our evaluation process, the dominant emotion is identified by the VQ-classifier. This obtained emotion label is then compared against the corresponding label associated with the reference input.
We use Wav2Vec 2.0  to measure finer affect qualities such as valence, arousal, and dominance. Furthermore, it is not enough to rely solely on quantitative metrics to measure finer nuances in affect. Therefore, we performed surveys to determine Mean Opinion Scores (MOS) and Emotion Perception Scores (EPS).
\subsubsection{Mel-Cepstral Distortion (MCD)}
Mel-Cepstral distortion serves as a quantitative metric, enabling the assessment of dissimilarities between two Mel-Spectrograms.
\subsubsection{Pearson Correlation Coefficient (PCC)}
Pitch is a significant factor influencing speech emotion. Pitch can be represented with fundamental frequency F0. Pearson Correlation Coefficient is a metric that allows us to determine the correlation between two pitch sequences.  
\subsubsection{Structural Similarity Index (SSIM)}
We used the structural similarity index as a metric to compare the similarity between ground truth and generated mel-spectrograms during the training time. SSIM was used as the indicator metric to determine the quality of the generated mel-spectrogram. 

\subsection{Qualitative Evaluations}
As part of the qualitative evaluation, users were asked to evaluate the quality of emotion in the synthesized speech based on the emotion from the reference speech. They were also asked to identify the emotion of the synthesized speech. 
\subsubsection{Mean Opinion Score (MOS)}
The Mean Opinion Score, or MOS, was calculated based on the quality of the  synthesized speech. The survey presented the users with three speech samples, the first being the neutral speech input prompt, the second being the reference speech and the third one being the generated speech with the target affect. The users were asked to rank the samples on a scale of 1 to 5, with 1 being the lowest in quality. There were 16 speech triplets that were presented to the users in four categories - English, with English reference, English, with Chinese reference; Chinese, with Chinese reference and Chinese, with English reference.
\subsubsection{Emotion Perception Score (EPS)}
For the computation of the emotion perception score, participants were presented with a task wherein they were required to select the dominant emotion conveyed by the synthesized speech from a set of five predefined options, namely, Sad, Happy, Angry, Neutral, and Surprised. The user-annotated emotion option was subsequently compared with the Speech Emotion Recognition (SER) label assigned to the reference speech.  

\section{Experiments}
The input mel-spectrograms are generated from raw audio samples in wav format. Torchaudio's mel-spectrogram transform function is used to convert the audio into the mel-spectrogram. The number of bins is set to 80, and the length of the fast Fourier transform is set to 2048. Window length and hop length are set to 1200 and 300, respectively. The input to the models is the log-normalized version of this mel-spectrogram, which was experimentally found to perform better than mel-spectrograms
\subsubsection{One-to-One Emotion Mapping (Same Speaker)}
This experiment tested the model's ability to translate emotions when the reference embeddings correspond to the same speaker.
The following table summarizes the performance of the model on both English and Chinese speech samples. MCD refers to Mel-Cepstral distortion. Lower values of MCD indicate higher quality of output. SSIM is the structural similarity between the ground truth and the generated output. SER is the emotion recognition score, and MOS is the mean opinion score. Higher values of SSIM, SER, and MOS indicate better quality of output.  
\begin{table}[h!]
    \centering
    
    \begin{tabular}{| c | c | c | c | c |}
       \hline
       \textbf{Emotion}  & \textbf{MCD} & \textbf{SSIM} & \textbf{SER} & \textbf{MOS}\\
       \hline
       Angry  & 5.16 & 0.69 & \textbf{0.76} & 3.79 $\pm$ 0.93\\
       Happy & 4.88 & 0.71 & 0.50 & 4.00 $\pm$ 1.15\\
       Sad & \textbf{4.42} & \textbf{0.72} & \textbf{0.76} & \textbf{4.11 $\pm$ 0.96}\\
       Surprised & 5.40 & 0.68 & 0.63 & 3.92 $\pm$ 1.20\\ 
       \hline
    \end{tabular}
    \caption{\textit{The table presents the performance of the model in generating synthetic speech with the dominant emotion in the English. }}
    \label{tab:eng_quant_perf}
\end{table}

\begin{table}[h!]
    \centering
    \begin{tabular}{| c | c | c | c | c | }
       \hline
       \textbf{Emotion}  & \textbf{MCD} & \textbf{SSIM} & \textbf{SER} & \textbf{MOS}\\
       \hline
       Angry  & 5.63 & 0.67 & 0.87 & 3.78 $\pm$ 1.12\\
       Happy & 5.52 & 0.68 & 0.61 & \textbf{3.95 $\pm$ 0.95}\\
       Sad & \textbf{4.05} & \textbf{0.74} & \textbf{0.89} & 3.78 $\pm$ 1.08\\
       Surprised & 5.65 & 0.66 & 0.71 & 3.68 $\pm$ 1.23 \\
       \hline
    \end{tabular}
    \caption{\textit{The table presents the performance of the model in generating synthetic speech with the dominant emotion in the Chinese Language}}
    \label{tab:chi_quant_perf}
\end{table}
In tables \ref{tab:eng_quant_perf} and \ref{tab:chi_quant_perf}, it can be observed that sadness is modeled better than all other emotions. Further investigations showed that this was due to the fact that happiness, surprise, and anger had very similar valence, arousal, and dominance values. In all of the experiments, the models performed slightly better on Chinese speech samples than on English ones. This could be because Chinese is a tonal language and has been shown to be denser tonally than English \cite{zhang2008acoustic}. 
\subsubsection{Same Language speaker Independent Emotion Translation}
In this setting, we tested the quality of generated emotions when both the reference speech and the input speech are in the same language, but not the same speaker. This aims to test the ability of the model to capture intonations and affect features accurately. 
 
\begin{table}[h!]
    \centering
    \begin{tabular}{| c | c | c | c | c |}
       \hline
       \textbf{Emotion}  & \textbf{MCD} & \textbf{SSIM} & \textbf{SER} \\
       \hline
       Angry  & 5.40 & 0.53 & \textbf{0.73}  \\
       Happy & 5.10 & 0.57 & 0.48  \\
       Sad & \textbf{4.61} & \textbf{0.62} & 0.72 \\
       Surprised & 5.49 & 0.53 & 0.61 \\ 
       \hline
    \end{tabular}
    \caption{\textit{The table presents the performance of the model in generating speaker-independent synthetic speech with the dominant emotion in the English Language}}
    \label{tab:eng_quant_perf_ind}
\end{table}

\begin{table}[h!]
    \centering
    \begin{tabular}{| c | c | c | c | c |}
       \hline
       \textbf{Emotion}  & \textbf{MCD} & \textbf{SSIM} & \textbf{SER} \\
       \hline
       Angry  & 5.81 & 0.51 & 0.83 \\
       Happy & 5.72 & 0.52 & 0.63 \\
       Sad & \textbf{4.28} & \textbf{0.67} & \textbf{0.85} \\
       Surprised & 6.10 & 0.50 & 0.65 \\ 
       \hline
    \end{tabular}
    \caption{\textit{The table presents the performance of the model in generating speaker-independent synthetic speech with the dominant emotion in the Chinese Language}}
    \label{tab:chi_quant_perf_ind}
\end{table}

It can be observed in Tables \ref{tab:eng_quant_perf_ind} and \ref{tab:chi_quant_perf_ind}, that when the model was conditioned on previously unseen embedding, the quality of the output with respect to the ground truth dropped slightly. This, however, is not an indication of the model's performance because the model generated the speech with a variation of the dominant emotion. Therefore, the output has to be different from the ground truth. The Mean Opinion score serves as a better metric to evaluate the quality of the output.

\subsubsection{Cross Language Speaker Independent Emotion Translation} In this setting, we tested the ability of the model to translate intonations from one language to another. Different languages employ varied styles of expressing emotions. Accurately translating the affect qualities across languages ensures the ability of the model to map variations in affect while exhibiting the same dominant emotion. The reference speech was in a different language from the input speech and had a different speaker and different linguistic content. 

\begin{table}[h!]
    \centering
    \begin{tabular}{| c | c | c | c | c | c |}
       \hline
       \textbf{Emotion}  & \textbf{MCD} & \textbf{SSIM} & \textbf{SER} & \textbf{EPS}\\
       \hline
       Angry  & 5.47 & 0.60 & \textbf{0.83} & 0.81\\
       Happy & 5.06 & 0.55 & 0.66 & 0.50\\
       Sad & \textbf{4.14} & \textbf{0.63} & 0.78 & \textbf{0.87}\\
       Surprised & 5.48 & 0.56 & 0.67 & 0.61\\ 
       \hline
    \end{tabular}
    \caption{\textit{The table presents the performance of the model in generating speaker-independent synthetic speech with the dominant emotion in either of the two languages, randomly picked}}
    \label{tab:chi_eng_quant_perf_dual}
\end{table}
In this setting, from table \ref{tab:chi_eng_quant_perf_dual}, a notable observation is the model's performance, even when conditioned on samples from a different language. This outcome demonstrates the model's ability to effectively capture and utilize information across distinct distributions. This attribute serves as a significant advantage of utilizing the vector quantized codebook. The generated embeddings consistently map to a known vector within the codebook, exhibiting the highest cosine similarity. Consequently, the generator model is always conditioned on a known affect vector, irrespective of the language, ensuring consistent and reliable affect representation in the synthesized speech across different linguistic contexts. 

\begin{figure*}[h!]
    \centering
    \includegraphics[width=1\textwidth, height=110px]{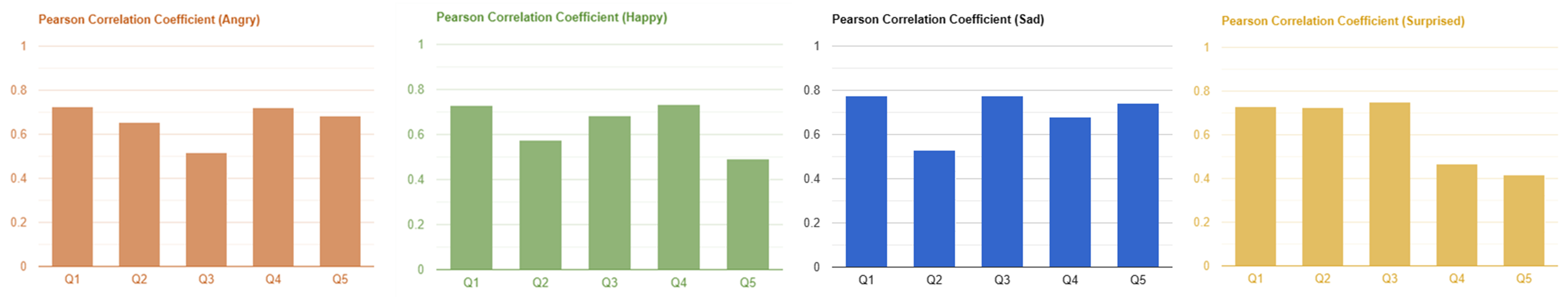}
    \caption{\textit{The graph shows trends in Pearson Correlation Coefficients across different emotions. It can be observed that the correlation varies with the ground truth, representing pure emotion. This indicates that the five embeddings represent varying levels of the same emotion}}
    \label{fig:pcc-plot}
\end{figure*}

Figure \ref{fig:pcc-plot} represents the plot of Pearson Correlation Coefficients across different emotions in the cross-language setting. It can be observed that these values have varying degrees of correlation with the ground truth speech sample, indicating that the linguistic content, speaker identity and speaking style remain consistent, but the affect qualities vary across the different vectors of the codebook.

\subsubsection{Translation accuracy by emotion}
This assessment serves two key purposes: firstly, to determine any potential biases towards specific emotions within the model, and secondly, to assess the overall separability of emotions from one another. By conducting this experiment, we aimed to identify correlations between emotional states and the limitations in separating similar but distinct emotional states. 
\begin{table}[h!]
    \centering
    \resizebox{8.5cm}{!}{
    \begin{tabular}{| c | c | c | c | c | c |}
        \hline
        \textbf{Emotion} &  \textbf{Angry} & \textbf{Happy} & \textbf{Neutral} & \textbf{Sad} & \textbf{Surprised}\\
        \hline
        Angry & 93.17\% & 3.76\% & 2\% & 0.43\% & 0.56\%\\
        Happy & 12.5\% & 76.7\% & 1.26\% & 1.06\% & 8.36\% \\
        Neutral & 0.76\% & 0.2\% & 96.96\% & 2.03\% & 0.03\% \\
        Sad & 0.6\% & 0.36\% & 2.3\% & 96.56\% & 0.16\% \\
        Surprised & 2\% & 5.1\% & 0.03\% & 0.33\% & 92.2\% \\
        \hline
    \end{tabular}
    }
    \caption{\textit{Emotion Embedding classification accuracy for English Speech Samples}}
    \label{tab:eng-acc}
\end{table}
\begin{table}[h!]
    \centering
    \resizebox{8.5cm}{!}{
    \begin{tabular}{| c | c | c | c | c | c |}
        \hline
        \textbf{Emotion} &  \textbf{Angry} & \textbf{Happy} & \textbf{Neutral} & \textbf{Sad} & \textbf{Surprised}\\
        \hline
        Angry & 96.8\% & 1.9\% &  0.96\%& 0\% & 0.33\% \\
        Happy & 7.4\% & 85.4\% & 0.36\% & 0.43 \% & 6.4\%\\
        Neutral & 0.3\% & 0.03\% & 98.4\% & 1.2\% &0\%\\
        Sad & 0.2\% & 0.03\% & 0.8\% & 98.6\% & 0.3\% \\
        Surprised & 1.1\% & 4.43\% & 0.43\% & 0.3\% & 93.7\% \\
        \hline
    \end{tabular}
    }
    \caption{\textit{Emotion Embedding classification accuracy for Chinese Speech Samples}}
    \label{tab:chi-acc}
\end{table}
It is apparent from tables \ref{tab:eng-acc} and \ref{tab:chi-acc} that the model performs slightly better on Chinese samples. Furthermore, happiness is most likely to be misclassified in both cases, with it being misclassified as either anger or surprise. From figure \ref{fig:v-a-d_plot}, it can be concluded that this is because the valence-arousal-dominance values of these emotions are very similar. Therefore, it is difficult to distinguish them from each other. 
\subsubsection{Skewness in Quantized Embeddings}
We conducted experiments to investigate potential biases within the model towards specific quantized embeddings over others. The aim was to ascertain if the model displayed preferences or imbalances in representing certain emotional nuances. Additionally, we explored the possibility of language dependencies within these quantized embeddings, assessing whether the emotion representations varied across different languages. 

\begin{table}[h!]
    \centering
    \resizebox{8.5cm}{!}{
    \begin{tabular}{|c|c|c|c|c|c|}
         \hline
         \textbf{Emotion} & \textbf{Q1} & \textbf{Q2} & \textbf{Q3} & \textbf{Q4} & \textbf{Q5}\\
         \hline
         Angry & 49.9\% &11.14\% & 0.075\% & 38.55\% & 0.265\% \\
         Happy & 29.6\% & 10.4\% & 16.09\% & 36.5\% & 7.2\% \\
         Neutral & 24.51\% & 0.30\% & 9.5\% & 38.4\% & 27.1\%\\
         Sad & 55.29\% & 0.03\% & 33.69\% & 10.69\% & 0.27\% \\
         Surprised & 16.6\% & 41.56\% & 37.53\% & 4.10\% & 0.18\%\\
         \hline
    \end{tabular}
    }
    \caption{\textit{Distribution of dominant emotions within the five quantized spaces in the ESD English speech corpus}}
    \label{tab:eng-emo-dist}
\end{table}

\begin{table}[h!]
    \centering
    \resizebox{8.5cm}{!}{
    \begin{tabular}{|c|c|c|c|c|c|}
         \hline
         \textbf{Emotion} & \textbf{Q1} & \textbf{Q2} & \textbf{Q3} & \textbf{Q4} & \textbf{Q5}\\
         \hline
         Angry & 41.93\% &5.83\% & 0.14\% & 51.45\% & 0.63\% \\
         Happy & 35.27\%\% & 6.41\% & 17.09\% & 35.59\% & 5.71\% \\
         Neutral & 28.73\% & 0.43\% & 5.98\% & 35.42\% & 29.41\%\\
         Sad & 50.93\% & 0.06\% & 43.07\% & 5.58\% & 0.33\% \\
         Surprised & 18.71\% & 33.12\% & 44.59\% & 3.49\% & 0.07\%\\
         \hline
    \end{tabular}
    }
    \caption{\textit{Distribution of dominant emotions within the five quantized spaces in the ESD Chinese speech corpus}}
    \label{tab:chi-emo-dist}
\end{table}
Tables \ref{tab:eng-emo-dist} and \ref{tab:chi-emo-dist} show that the distribution for English and Chinese speech samples is quite similar, indicating that the mapping is independent of the language. Some of the embeddings, such as Angry-Q3 and Neutral-Q2, are rarely chosen. Upon further investigation, it was found that these embeddings contained information corresponding to the emotions such as fear, contempt, and disgust, which were absent in the ESD dataset, but were present in the MEAD dataset. 
\subsubsection{Quantifying Valence-Arousal-Dominance}
A key contribution of our work is to represent finer nuances in affect features beyond the simple representation of dominant emotion. To verify that the embeddings exhibit variations in valence, arousal, and dominance, we use the finetuned wav2vec 2.0-dimensional emotion model \cite{wagner2023dawn} to calculate these attributes from the generated mel-spectrogram. 
\begin{figure}[h]
    \centering
    \includegraphics[width=7cm]{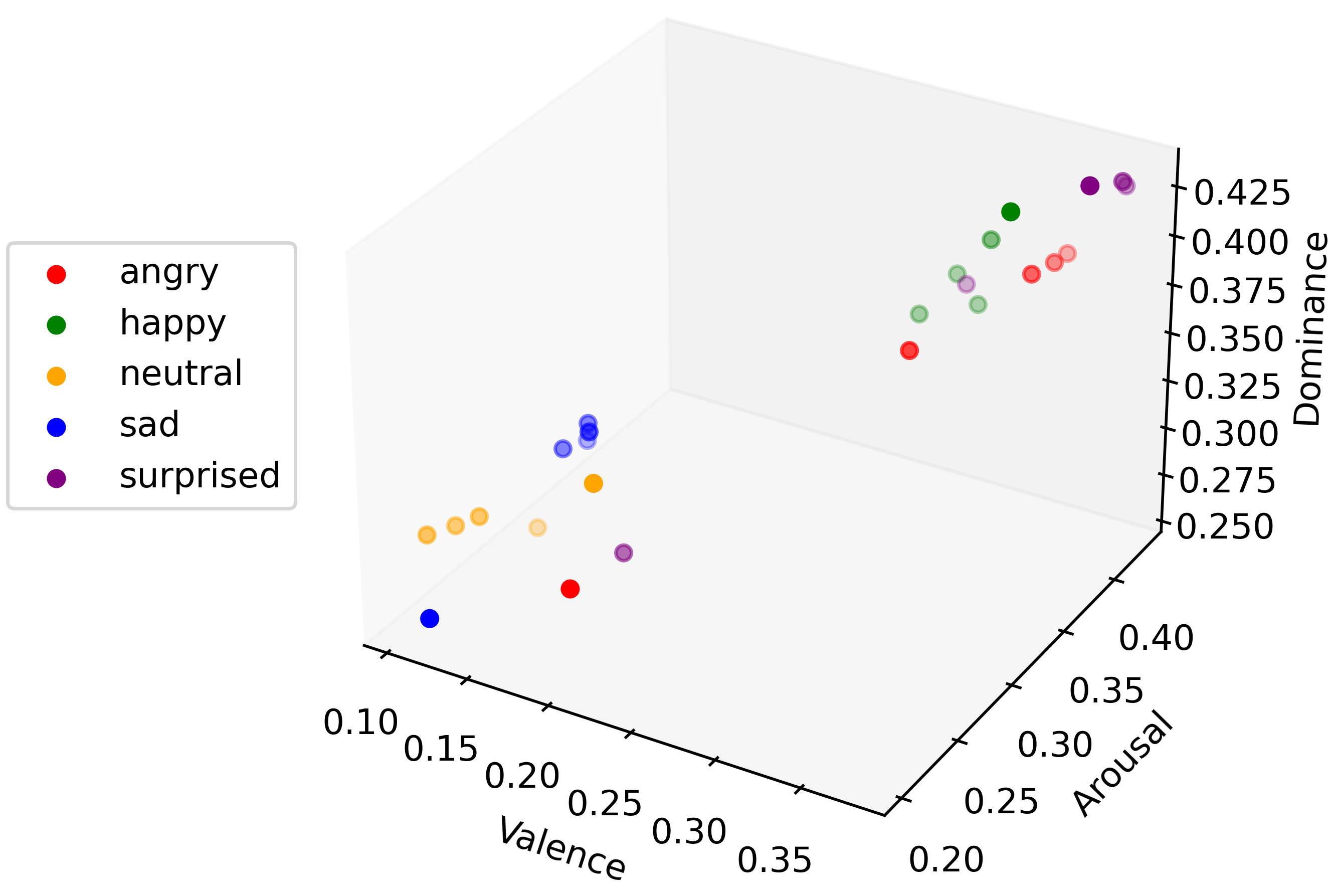}
    \caption{\textit{Scatter plot of average valence-arousal-dominance values of 200 generated audio samples in both English and Chinese}}
    \label{fig:v-a-d_plot}
\end{figure}\\

\begin{figure*}[h]
\centering
\begin{minipage}{.5\textwidth}
  \centering
  \includegraphics[width=6.5cm]{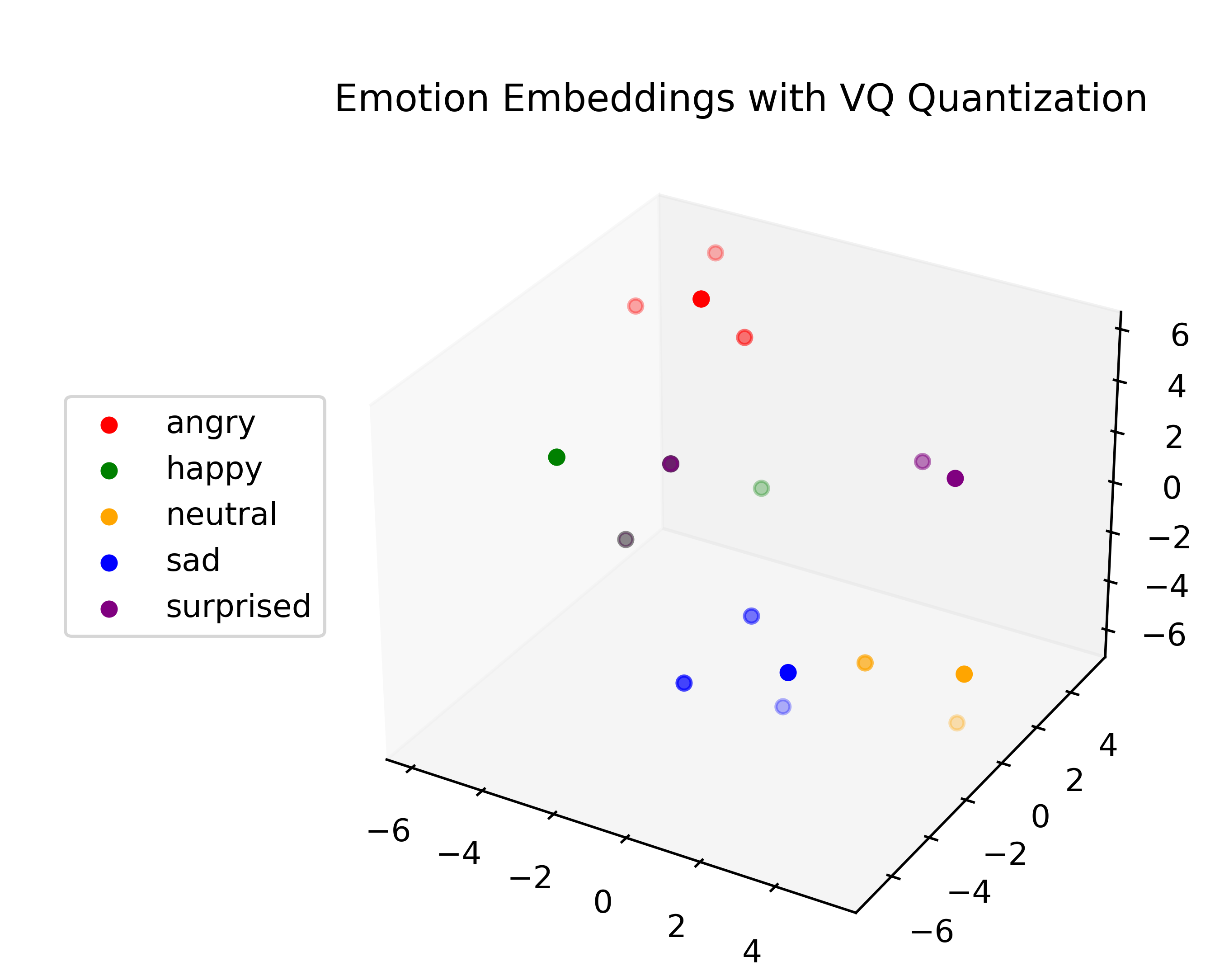}
  \captionof{figure}{\textit{Visualization of quantized embeddings \newline in 3D space through tSNE. It can be observed that similar \newline emotions are grouped together}}
  \label{fig:quantized-cluster}
\end{minipage}%
\begin{minipage}{.5\textwidth}
  \centering
  \includegraphics[width=6.5cm]{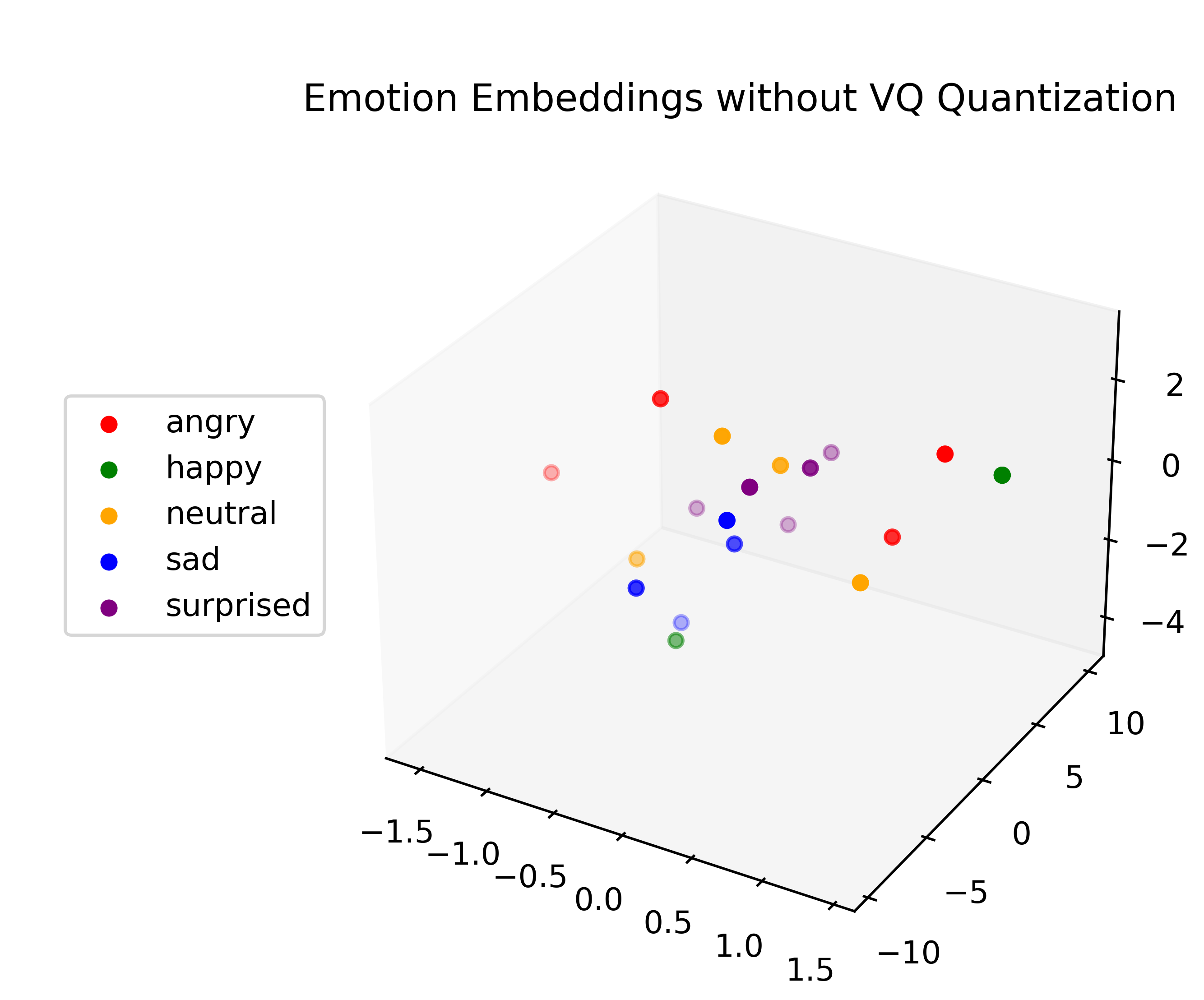}
  \captionof{figure}{\textit{Visualization of the embeddings generated by the classifier model without the VQ Codebook. It can be observed that there's no clustering of emotions}}
  \label{fig:no-quantized-scatter}
\end{minipage}
\end{figure*}

Figure \ref{fig:v-a-d_plot} represents a scatter plot of average valence-arousal-dominance values generated on 200 samples of audio. Each point represents the audio conditioned on one of the five quantized vectors from the VQ Codebook. It can be observed that the dominant emotions exhibit similar values of valence and dominance but vary in arousal. Sadness has the lowest valence, while surprise has the highest value of valence. 

\section{Ablation Studies}
\subsection{Effect of Vector Quantization}
To understand the effect of the VQ Quantization codebook, we built a classifier, but without the codebook.
We use t-distributed Stochastic Neighbor Embedding (tSNE) visualizations to demonstrate the effectiveness of the quantized emotion space. Notably, we observe that the embeddings corresponding to dominant emotions are grouped together, indicating successful clustering while simultaneously diverging in the direction of the less dominant emotion. This divergence signifies the varying intensity of the dominant emotion, showcasing the model's capability to represent emotional expressions with finer granularity.\\

\begin{figure}[h]
    \centering
    \includegraphics[width=7cm]{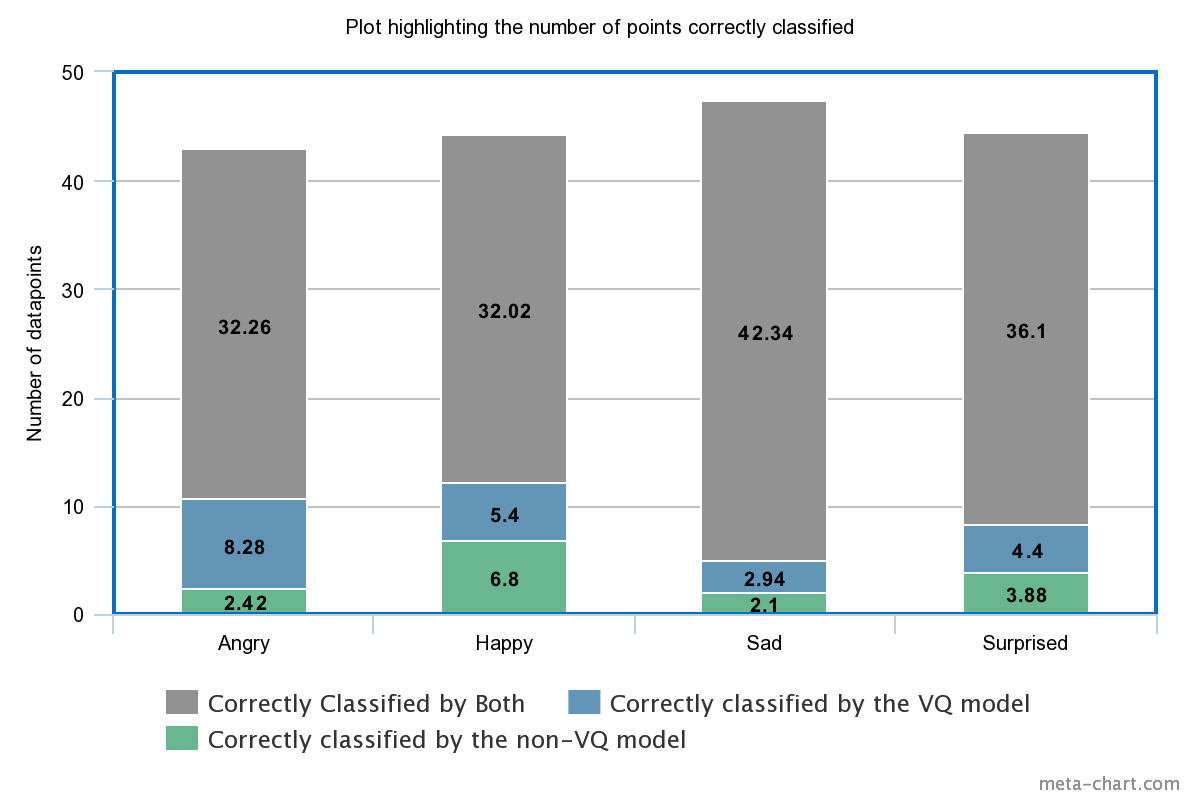}
    \caption{\textit{The graph highlights the percentage of points correctly classified by the two models}}
    \label{fig:misclass}
\end{figure}
\begin{table}[h!]
    \centering
    \begin{tabular}{|c|c|}
         \hline
         \textbf{Emotion} & \textbf{p-value}\\
         \hline
         Angry & 0.005 \\
         Happy & 0.2  \\
         Neutral & 0.014\\
         Sad & 0.001\\
         Surprised & 0.001 \\
         \hline
    \end{tabular}
    \caption{\textit{One-Tailed T-test scores comparing the performance of the VQ-Classifier model against the non-VQ model over 50 trials.}}
    \label{tab:ttest}
\end{table}

From Figure \ref{fig:quantized-cluster}, it is apparent that neutral and sadness are grouped closer, while anger, happiness, and surprise are higher. Happiness and Surprise are somewhat entangled with each other, indicating the difficulty in distinguishing the two. Figure \ref{fig:no-quantized-scatter}, however, exhibits very little clustering in terms of emotion because it represents the embeddings generated without the vector quantization. In this scenario, each embedding independently contains affect information. 
Figure \ref{fig:misclass} highlights the percentage of audio correctly classified by the two models, with gray indicating the ones correctly classified by both models. It can be observed that the VQ codebook model correctly classifies a larger percentage of the dataset than the non-VQ model. Table \ref{tab:ttest} shows a one-tailed t-test that VQ-classifier outperforms the non-VQ-classifier for all emotions except happiness. 

\subsection{Effect of Spectral Convolution Layers}
The use of global convolution in the Spectral domain, also termed as \textit{neural operator},  has shown promise in vision tasks \cite{guibas2021adaptive}. More recently, it was shown by \cite{shchekotov2022ffc} that the neural operator outperformed convolution in speech generation models. We trained two versions of the generator model, one with a spectral convolution block and one with only a regular convolution block. To test the difference in the obtained results, we performed the Wilcoxon Signed Rank Test, testing whether the outputs of the neural operator model had higher SSIM than the outputs of the model without the neural operator. The metric used for comparison was 50 instances of average SSIM of 50 randomly sampled data points from the test set. The p-value was 0.00224, with a v value of 923.0, indicating that the neural operator improved the model.

\begin{table}[h!]
    \centering
    \begin{tabular}{| c | c | c |}
       \hline
       \textbf{Emotion}  & \textbf{MCD}  & \textbf{MOS}\\
       \hline
       Mixed Emotions  & 6.62 & 3.45 $\pm$ 0.12\\
       EmoDiff  & 5.72 & 4.14 $\pm$ 0.10\\
       GradTTS  & 5.78 & 4.13 $\pm$ 0.10\\
       AffectEcho & \textbf{5.47} & \textbf{4.16 $\pm$ 0.91}\\ 
       \hline
    \end{tabular}
    \caption{\textit{Comparisons against other speech generation architectures. Lower MCD indicates better results. Higher MOS indicates better quality output}}
    \label{tab:mod_comps}
\end{table}

\section{Discussion and Conclusion}
In our work, we show that vector quantized embeddings are, in a broad sense, interpretable; however, the individual features of the 64-feature vector representing the quantized emotion cannot be tuned manually to translate the emotion. The use of quantized vectors ensures that out-of-distribution speech samples are still mapped to the nearest known embeddings, preserving the robustness and generalization capacity of the emotion representation.\\
The model was shown to perform well in cross-language settings and accurately mapped similar emotions together, across languages. In table \ref{tab:mod_comps}, it can be observed that AffectEcho has an overall lower MCD score. Furthermore, while the other models generate emotional speech with good accuracy, they fail to incorporate affective features from the human speaker, with whom they are interacting. Our model aims to bridge this gap and showcase stronger human-AI interaction capabilities. \\
For further improvement in this architecture, the following directions should be explored 
\begin{itemize}
    \item The use of consistency models to sample speeches from emotion prompts, allowing even greater variations in emotional manifestations
    \item Adapting this architecture for longer monologues with multiple emotions
    \item integrating this with text-to-speech models to build end-to-end dialogue systems. 
\end{itemize}
\bibliography{aaai24}

\end{document}